\documentclass[journal]{IEEEtran}
\IEEEoverridecommandlockouts

\usepackage{amsmath,amssymb,amsfonts}
\usepackage{algorithmic}
\usepackage{graphicx}
\usepackage{textcomp}
\usepackage{xcolor}
\usepackage{amssymb}
\usepackage{algorithmic}            
\usepackage{algorithm}
\usepackage{lipsum} 

\begin{document}
    \title{\huge Study of Iterative Detection and Decoding for Multiuser Systems and MMSE Refinements with Active or Passive RIS}

\author{Roberto~C.~G.~Porto and
        Rodrigo~C.~de~Lamare,~\IEEEmembership{Senior Member,~IEEE}\vspace{-1.75em}  
        
\thanks{The authors are with the Centre for Telecommunications Studies,
Pontifical Catholic University of Rio de Janeiro, Rio de Janeiro 22453-900,
Brazil (e-mail: camara@ime.eb.br; delamare@puc-rio.br).

This work was supported by CNPq, CAPES, FAPERJ e FAPESP.}}

\maketitle

    \begin{abstract}
An iterative detection and decoding (IDD) scheme is proposed for multiuser multiple-antenna systems assisted by an active or a passive Reconfigurable Intelligent Surface (RIS). The proposed approach features an IDD strategy that incorporates Low-Density Parity-Check (LDPC) codes, RIS processing with refinements of soft information in the form of log likelihood ratios (LLRs) and truncation. Specifically, a minimum mean square error (MMSE) receive filter is used for refinement of LLRs and truncation at the RIS, and for soft interference cancellation at the receiver. 
An analysis of the proposed MMSE refinement is also devised along with a study of the computational complexity of the proposed and existing schemes. Simulation results demonstrate significant improvements in system capacity and bit error rate in the presence of block-fading channels.
\end{abstract}


    \vspace{-1em}
\section{Introduction}

The escalating demand for higher data rates, enhanced coverage and reduced power consumption in wireless communications requires advanced technologies. A Reconfigurable Intelligent Surface (RIS) exhibits significant potential for optimizing wireless networks and are expected to be incorporated in the sixth-generation (6G) of wireless communication systems \cite{rista} that are equipped with multiple antenna systems \cite{mmimo,wence}. Unlike conventional methods that alter transmitted signals, a RIS modifies the characteristics of the equivalent channel between users and access points \cite{9140329}. A RIS has the potential to extend coverage \cite{9359653}, reduce power consumption \cite{9548940}, and increase channel capacity \cite{9998527,9913356,9110912}.

The work in \cite{9452133} studied RIS-assisted wireless systems, where the data from the transmitter are encoded by Low-Density Parity-Check (LDPC) codes. The performance evaluation employs bit-error rate (BER) with results from simulations compared against analytical bounds. Moreover, the works in \cite{1494998, 8240730, 7105928} consider iterative detection and decoding (IDD) schemes \cite{spa,mfsic,mbdf,bfidd} using LDPC codes for multiple-antenna systems under block-fading channels. Additionally, \cite{9998527} introduces an active RIS as a solution to mitigate the limitations posed by a passive RIS due to the multiplicative fading effect.

Several studies have investigated uplink systems assisted by RIS \cite{9548940}, \cite{electronics13050969} - \cite{9966649}, each employing distinct optimization techniques, typically using SINR as the objective function. Among these, only one study utilized closed-form expressions (CE) \cite{9966649}. However, none have examined systems that integrate both RIS and iterative processing techniques. In this context, IDD schemes perform the exchange of soft information in the form of log-likelihood ratios (LLRs), between the detector and decoder. Previous research on multiple-antenna systems and IDD schemes with block-fading channels and LDPC codes has demonstrated performance improvements proportional to the number of information exchanges \cite{8240730, 7105928}. This study considers a RIS-aided multiuser multi-input single-output (MU-MISO) system in typical 3GPP wireless scenarios \cite{access2010further}. 

In this work, we present an IDD scheme for multiuser MISO systems assisted by an active or a passive RIS \cite{risidd}. The proposed IDD approach employs LDPC codes, RIS processing with the MMSE refinement of LLRs and truncation. To the best of our knowledge, this is the first work to integrate IDD with a RIS while utilizing MMSE processing for the exchange and refinement of LLRs, and truncation. In particular, an MMSE receive filter is used for refinement of LLRs and truncation at the RIS, and for soft interference cancellation at the receiver. An analysis of the proposed MMSE refinement is also devised along with a study of the computational complexity of the proposed and existing schemes. Numerical results show significant improvements in system capacity and bit error rate in the presence of block-fading channels.

The rest of this paper is organized as follows. In Section II we describe the system model. Section III describes the derivation of the proposed IDD scheme assisted by passive or active RIS. Section IV presents an analysis of the proposed MMSE refinement and the computation complexity of competing schemes. Section V discusses the simulation results and Section VI concludes this paper.

\textit{Notations:} Bold capital letters indicate matrices while vectors
are in bold lowercase. $\mathbf{I}_n$ denotes $n \times n$ identity matrix. Furthermore, diag($\mathbf{z}$) is a diagonal matrix consisting solely of the elements extracted from vector $\mathbf{z}$.

    \begin{figure*}
\vspace{-3em}
    \centerline{\includegraphics[width=.85\textwidth]{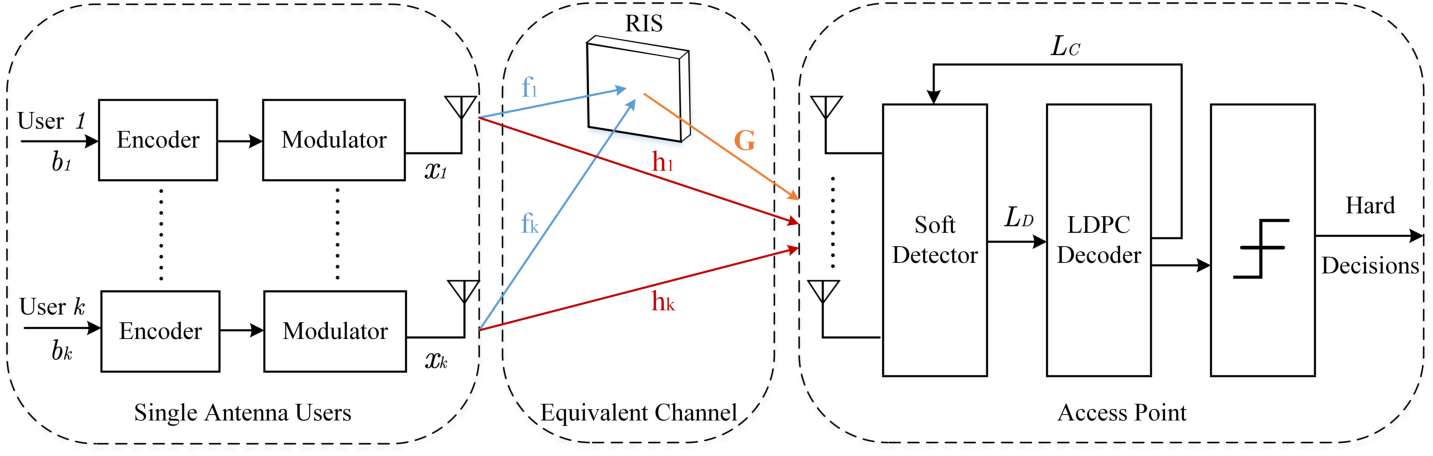}}
    \vspace{-0.725em}
    \caption{System model of an IDD multiuser multiple-antenna system.}
    \label{fig01}
    \vspace{-1em}
\end{figure*}
\section{System Model}

A single-cell MU-MISO system featuring a multiple-antenna access point (AP) assisted by a RIS is considered, as illustrated in Fig. \ref{fig01}. In this configuration, the AP is equipped with $M$ antennas, supporting $K$ users, each equipped with a single antenna. Initially, each user’s information symbols are encoded via individual LDPC channel encoders and subsequently modulated to $x_k$ employing a QPSK modulation scheme. The transmit symbols $x_k$ have zero mean and share the same energy, with $E[|x_k|^2] = \sigma^2_x$. These modulated symbols are then transmitted over block-fading channels.

The RIS is assumed to incorporate $N$ reflecting elements and their reflection coefficients are modeled as a complex vector $\boldsymbol{\varphi} \triangleq [p_1e^{j\theta_1}, \dots, p_ne^{j\theta_N}]^T$, where $\theta_n$ represents the phase shift of the $n$th unit, and $p_n \in \mathbb{R}_+$ denotes the gain factor of the  $n$th unit, for all $n \in \mathcal{N} \triangleq \{ 1, 2, \dots, N\}$. This expression allows for the representation of both passive and active RIS systems. The distinction between the two lies in the fact that in the passive system, $p_n = 1$, for all $n$.

The signal model of an $N$-element RIS is given by a diagonal phase shift matrix $\Phi \in {\mathbb{C}^{N\times N}}$ defined as $\mathbf{\Phi} \triangleq \text{diag}(\boldsymbol{\varphi})$. We can represent the equivalent channel from the AP to user $k$, which includes both the direct link and the reflected link, denoted as $\mathbf{\bar{h}_k} \in {\mathbb{C}^{M \times 1}}$, by 
\begin{equation}
    \mathbf{\bar{h}_k} = \mathbf{h_k} + \mathbf{G}\mathbf{\Phi}\mathbf{f_k},
    \label{heff}
\end{equation}
where $\mathbf{h_k} \in \mathbb{C}^{M \times 1}$, $\mathbf{G} \in \mathbb{C}^{M \times N}$, and $\mathbf{f_k} \in \mathbb{C}^{N \times 1}$ represent the communication links from the AP to the user $k$, from the AP to the RIS, and from the RIS to the user $k$, respectively. 

An estimate $\hat{x}_k$ of the transmitted symbol of the $k$th user without soft interference cancellation (SIC) is obtained by applying a linear receive filter $\mathbf{w_k}$ to the received signal $\mathbf{y}$:
\begin{equation}
    \hat{x}_k = \mathbf{w_k}^H\mathbf{y} = \mathbf{w_k}^H\left(\sum_{i=1}^K\mathbf{\bar{h}_k}x_k + \mathbf{G}\mathbf{\Phi}\mathbf{n_v} +\mathbf{n_s}\right),
    \label{detection_estimate_1}
\end{equation}
where the vector $\mathbf{\Phi}\mathbf{n_v}$ denotes the noise introduced and amplified by the active RIS , as described in \cite{9998527}. This noise is characterized as 
$\mathbf{n_v} \sim \mathcal{CN}(\mathbf{0_M}, \sigma_v^2 \mathbf{I_N})$. The amplified noise is subsequently multiplied by the communication links from the RIS and the AP, represented by $\mathbf{G}$. For a passive RIS, $\sigma_v$ is set to zero. The vector $\mathbf{n_s} \sim \mathcal{CN}(\mathbf{0_M}, {\sigma_s^2} \mathbf{I_M})$ represents the static noise. Since the matrix $\mathbf{\Phi}$ is diagonal, the symbol estimate in (\ref{detection_estimate_1}) can also be written in terms of $\boldsymbol{\varphi}$ as given by
\begin{equation}
    \hat{x}_k = \mathbf{w_k}^H\left(\sum_{i=1}^K\mathbf{h_k}x_k + \sum_{i=1}^K\mathbf{A_k}\boldsymbol{\varphi}x_k + \mathbf{B}\boldsymbol{\varphi} +\mathbf{n_s}\right),
    \label{detection_estimate_2}
\end{equation}
where $\mathbf{B}=\mathbf{G}\text{diag}(\mathbf{n_v})$ and $\mathbf{A_k} = \mathbf{G} \text{diag}(\mathbf{f_k})$.
    \section{Proposed Iterative Detection and Decoding}

Motivated by earlier research on IDD schemes \cite{1494998,8240730,spa,mfsic,bfidd,dfcc,jidf,sjidf,1bitidd,mbthp,did,rrber,dynovs,listmtc,lrcc,detmtc,msgamp1,msgamp2,arqidd,llridd,comp,oclidd}, an IDD scheme with a soft detector and LDPC decoding is devised to enhance the performance of the system. The soft detector incorporates extrinsic information provided by the LDPC decoder ($\mathbf{L_C}$), and the LDPC decoder incorporates soft information provided by the MIMO detector ($\mathbf{L_D}$).

The incorporation of RIS introduces challenges to the design of an uplink receive filter, primarily due to its dependence on the reflection coefficient vector ($\boldsymbol{\varphi}$). The optimization of $\boldsymbol{\varphi}$ requires information from the receive filter. To address this issue, we employ a scheme based on alternating optimization (AO). To optimize the reflection coefficient vector, it is computed before the IDD iterative loop. This decision is driven by the inefficiency of using alternating optimization within the IDD iteration loop to compute both $\mathbf{w_k}$ and $\boldsymbol{\varphi}$. Additionally, the required information exchange between the RIS and the AP would be impractical. Therefore, a criterion aimed at maximizing the gains from the IDD is chosen, as will be presented in Section IV.

\subsection{Design of Reflection Parameters}
\label{sec:designRIS}

The symbol estimate without SIC from (\ref{detection_estimate_2}) is utilized for the design of the reflection parameters. The problem formulation begins by fixing the value of the reception filter and seeking the vector $\boldsymbol{\varphi}$ that minimizes $E[||\mathbf{x} - \hat{\mathbf{x}}||^2_2]$, subject to constraints in accordance with the RIS operating principle (passive or active). Substituting the vector form of (\ref{detection_estimate_2}) that involves all user symbols leads to the following problem formulation:
\vspace{-0.25em}
\begin{equation}
\underset{\boldsymbol{\varphi}}{{\text{minimize}}} \quad 
    E\left[||\mathbf{x}-\mathbf{W}(\sum_{i=1}^K \mathbf{\bar{h}_i}{x}_i +\mathbf{B}\boldsymbol{\varphi} + \mathbf{n_s})||^2_2\right]
    \label{eq:obj_f}
\end{equation} 
\vspace{-1em}
\begin{align} 
 {\text{subject to}}\quad & \text{(RIS - active)} & \text{or} & ~~~~\text{(RIS - passive)} 
\nonumber \\ & \text{P}_\text{A} +||\boldsymbol{\varphi}||^2\sigma_v^2\leq \text{P}_\text{RIS}  &  & ~~~~~|[\boldsymbol{\varphi}]_n|=1
\nonumber \\ &  \text{P}_\text{A} = \sum^K_{i=1} ||\mathbf{\Phi}\mathbf{f}_i||^2\sigma_x^2 & & ~~~~~ \sigma_v^2=0
\\ & p_n \geq 1 & &
\nonumber 
\end{align}
where $\mathbf {W} \triangleq [\mathbf{w_1}, \dots, \mathbf{w_K}]^H $ represents the reception filter matrix and $\text{P}_\text{RIS}$ is the reflection power of the active RIS.

While the norm consistently exhibits convex behavior, it is noteworthy that the restriction on the diagonal elements of $\boldsymbol{\varphi}$ does not form a convex set. To address this, a relaxation on all constraints is employed. By computing the derivative of the objective function with respect to $\boldsymbol{\varphi}$ and setting the result to zero, the following expression is obtained:
\begin{equation} 
\boldsymbol{\varphi}_o = \left[\boldsymbol{\beta}+\frac{\sigma_v^2}{\sigma_x^2}(\mathbf{WG})^H(\mathbf{WG})\right]^{-1}\boldsymbol{\Psi},
\label{eq:phi}
\end{equation}
where {\footnotesize$\boldsymbol{\beta} = \sum^{K}_{i}\mathbf{(WA_i)^H(WA_i)}$, $\boldsymbol{\Psi} = \sum_i^K\mathbf{(WA_i)^H(e_i-W\mathbf{\bar{h}_i})}$} and  $\mathbf{e_i}$ is a column vector with all zeros, except for a one in the ith element.

Given the relaxation of the constraint, it becomes necessary to truncate the computed value for the active RIS as follows
\begin{equation} 
    \boldsymbol{\varphi_t}^\text{active} =  \boldsymbol{\varphi_o} \left( \frac{\text{P}_\text{RIS}}{ \sum^K_{i=1} ||\mathbf{\Phi}\mathbf{f}_i||^2\sigma_x^2 +||\boldsymbol{\varphi}||^2\sigma_v^2}\right)^\frac{1}{2}.
\end{equation}
This truncation involves multiplying $\boldsymbol{\varphi}_o$ by a factor to enforce the use of the available power $P_\text{RIS}$. Increasing the power does not affect the terms reflected by the RIS or the dynamic noise in the SNIR composition; however, it is inversely proportional to the direct links of the users and the static noise. In a scenario with a considerable number of users, if the $P_\text{RIS}$ increases, the interference of the direct links of the other users and the static noise at the receiver will decrease. 
Note that due to the convex nature of the objective function in the context of QPSK modulation, it is guaranteed that this approach yields the global optimal solution for the problem.

For the passive RIS, the suboptimal $\boldsymbol{\varphi_t}^\text{passive}$  is found by solving the following problem: 
\begin{align*} 
\boldsymbol{\varphi}_t = \underset{\boldsymbol{\varphi}_t }{{\text{min}}} 
\quad ||\boldsymbol{\varphi}_o-\boldsymbol{\varphi}_t||^2
\\
\text{subject to} \quad |[\boldsymbol{\varphi}_t]_n| = 1 \quad \forall n,
\end{align*}
expanding the terms of the objective function yields the expression
\begin{equation}
||\boldsymbol{\varphi}_o-\boldsymbol{\varphi}_t||^2 = 
||\boldsymbol{\varphi}_o||^2 -2\text{Re}[\boldsymbol{\varphi}_t^H\boldsymbol{\varphi}_o] + ||\boldsymbol{\varphi}_t||^2.
\end{equation}

Since $||\boldsymbol{\varphi}_t||^2=n$ due to the constraint $|[\boldsymbol{\varphi}]_n|=1$, the problem can be redefined as maximizing the term $\text{Re}[\boldsymbol{\varphi}_t^H\boldsymbol{\varphi}_o]$. Rewriting $\text{Re}[\boldsymbol{\varphi}_t^H\boldsymbol{\varphi}_o]$ as a summation yields
\begin{equation}
    \text{Re}[\boldsymbol{\varphi}_t^H\boldsymbol{\varphi}_o] = \sum_{i=1}^N \text{Re}([\boldsymbol{\varphi}_t^*]_i[\boldsymbol{\varphi}_o]_i)
    \label{eq:truc_pass}
\end{equation}
Expressing $[{\varphi}_o]_i$ on polar form as $[{\varphi}_o]_i = |[{\varphi}_o]_i|e^{j\measuredangle [\varphi_o]_i}$, this expression is maximized when the phases of $[{\varphi}_t]_i$ are aligned with those of $[{\varphi}_o]_i$. Thus, the suboptimal $\boldsymbol{\varphi_t}^\text{passive}$ is given by
\begin{equation}
    \boldsymbol{\varphi_t}^\text{passive} = \frac{[{\varphi}_o]_i}{|[{\varphi}_o]_i|} = e^{j\measuredangle (\boldsymbol{\varphi_0})}.
\end{equation}

\subsection{Iterative Detection and LDPC Decoding}
In the SIC detector \cite{1494998}, the received vector $\textbf{y}$ is processed through demapping via a log-likelihood ratio calculation for each bit among the coded bits transmitted by the users, represented by $\mathbf {x} \triangleq [x_1, \dots, x_K]^T$. Therefore, the extrinsic LLR value $L_D$ for the $i$-th code bit $b_i$ is computed as follows:
\begin{equation}
    L_{D}(b_{i})= \log\frac{\sum\nolimits_{{\bf x} \in {\cal X}_{i}^{+1}}P({\bf y}\vert {\bf x}, {\bf \bar{H}})P({\bf x})} {\sum\nolimits_{{\bf x} \in {\cal X}_{i}^{-1}}P({\bf y}\vert {\bf x},{\bf \bar{H}})P({\bf x})}-L_{C}(b_{i})
    \label{eq:ldlc}
\end{equation}

Inspired by prior work on IDD schemes \cite{1494998} and \cite{8240730}, the soft estimate of the $k$th transmitted symbol is firstly calculated based on the $\mathbf{L_c}$ (extrinsic LLR) provided by the channel decoder from a previous stage:
\begin{equation*} 
\tilde {x}_{k}=\sum _{x\in \mathcal {A}}x\text {Pr}(x_{k}=x)=\sum _{x\in \mathcal {A}}x\left ({\prod _{l=1}^{M_{c}}\left [{1+\text {exp}(-x^{l}L_{c}^{l})}\right]^{-1}}\right), 
\end{equation*}
where $\mathcal {A}$ is the complex constellation set with $2^{M_c}$ possible points. The symbol $x^l$ corresponds to the value $(+1, -1)$ of the $l$ th bit of symbol $x$ . 

A symbol estimate uses SIC, where the value of $\boldsymbol{\varphi}$ is fixed and $\mathbf{w_k}$ is chosen to minimize the mean square error (MSE) between the transmitted symbol $x_k$ and the filter output:
{\begin{equation} 
    \mathbf {w}_{k}=\arg \min _{ \tilde{\mathbf{w}}_{k}} E\left [{\left \vert{ x_{k}-\tilde{\mathbf{w}}_{k}^{H}\mathbf {y}_k}\right \vert ^{2}}\right]. 
\end{equation}}
It can be shown that the solution is given by
{\begin{equation}
    \mathbf{w_k} = \left(\frac{(\text{G}_\text{loss}\text{P}_\text{RIS}\sigma^2_v+\sigma^2_s)}{\sigma^2_x}\mathbf{I_{n_r}} + \mathbf{\bar{H} \Delta_k \bar{H}}^H \right)^{-1}\mathbf{\bar{h}_k},
    \label{eq:w}
\end{equation}
}where { $\text{G}_\text{loss}$ is the average path loss between the AP and the RIS, $\mathbf {\bar{H}} \triangleq [\mathbf{\bar{h}_1}, \dots, \mathbf{\bar{h}_K}]^H$ is the equivalent channel and} the covariance matrix $\mathbf{\Delta_k}$  is
{\begin{equation}
    \mathbf{\Delta_k} = \text{diag}\left[\frac{\sigma^2_{x_{1}}}{\sigma^2_x}\dots \frac{\sigma^2_{x_{k-1}}}{\sigma^2_x}, 1, \frac{\sigma^2_{x_{k+1}}}{\sigma^2_x},\dots,\frac{\sigma^2_{\sigma^2_{K}}}{\sigma^2_x}  \right],
\end{equation}}
where $\sigma^2_{x_{i}}$ is the variance of the $i$th user computed as:
{\begin{equation}
\sigma_{x_{i}}^{2}=\sum\limits_{x\in {\cal A}}\vert x-\bar{x} _{i}\vert ^{2}P(x_{i}=x). 
\end{equation}}

The pseudo-code of the proposed RIS-assisted IDD scheme is described in Algorithm 1.

\vspace{-0.5em}

\begin{algorithm}[H]
\footnotesize
    \label{algor1}
    \begin{algorithmic}[1]
        \caption{Proposed RIS-assisted IDD Scheme}\label{alg:cap}
        \STATE \textbf{Input:} Channels: $\mathbf{G, F, H}$ and received signal: $\mathbf{y}$.
        \STATE \textbf{Output:} $\mathbf{W}$, $\boldsymbol{\varphi_0}$
        \STATE Randomly initialize $\boldsymbol{\varphi}$, then compute $\mathbf{W}$ and $\boldsymbol{\varphi_0}$ using AO and (\ref{eq:phi});

        \FOR{$i=1$ to idd.iterations}
                        \STATE \textbf{Detection Scheme - SIC}
                        \STATE {Obtain the receiver filter $\mathbf{W}$ with (\ref{eq:w}).}
                        \STATE {Obtain the Extrisic bit LLRs $\mathbf{L_D}$.}
                \STATE \textbf{Proceed with LDPC decoding.}
                \STATE Compute the LLR value $\mathbf{L_C}$. 
        \ENDFOR
    \end{algorithmic}
\end{algorithm}
\vspace{-0.5cm}

\section{Analysis}
\label{sec:analysis}

In this section, we carry out an analysis of the proposed LLR refinement with MMSE processing and detail the computational cost of the proposed IDD scheme.

\subsection{LLR Refinement with MMSE Processing}

To enhance the performance of the IDD scheme, $\boldsymbol{\varphi}$ can be selected to amplify the magnitude of $L_D$ in (\ref{eq:ldlc}), thereby aiding the decoder's decision-making process \cite{7105928}. Analyzing for QPSK modulation, two $L_D$ values need to be computed for each received symbol, designated as $L_D^i$ and $L_D^{ii}$ \cite{1494998}: 
{
\begin{equation}
L_{D}^{i} = \ln{
\frac{P(\hat{x}_k \vert x_{10})P(x_{10}) + P(\hat{x}_k \vert x_{11})P(x_{11})}
{P(\hat{x}_k \vert x_{00})P(x_{00}) + P(\hat{x}_k \vert x_{01})P(x_{01})}}
- L_{C}^i
\label{eq:ldi1}
\end{equation}

\begin{equation}
L_{D}^{ii} = \ln{
\frac{P(\hat{x}k \vert x_{01})P(x_{01}) + P(\hat{x}k \vert x_{11})P(x_{11})}
{P(\hat{x}_k \vert x_{00})P(x_{00}) + P(\hat{x}_k \vert x_{10})P(x_{10})}}
- L_{C}^{ii}.
\label{eq:ldi2}
\end{equation}
}

The likelihood function approximation used is
$P(\hat{x}_{k} \vert x) \simeq \frac{1}{\pi \eta_k^{2}} \exp\left( -\frac{1}{\eta_{k}^{2}} \left\vert \hat{x}_{k} - \mu_{k}x \right\vert^{2} \right)$, where
$\mu_{k} \overset{\Delta}{=} E\left[ \hat{x}_{k}^{\text{sic}} \vert x \right] = \mathbf{w}^H_{k} \mathbf{h}_{k}$ and
$\eta_{k} \overset{\Delta}{=} \text{var}\left[ \hat{x}_{k}^{\text{sic}} \vert x \right] = {\sigma^2_x} (\mu_{k} - \mu_{k}^2)$
as described in \cite{1494998}. Gray coding is applied with the mapping $[x_{00}, x_{01}, x_{10}, x_{11}] = \sigma_x [-1+i, 1+i, -1-i, 1-i]$. 

{These values are substituted into (\ref{eq:ldi1}), and the terms are rearranged to yield the following expression:
\begin{align} 
    L_{D}^{i}= &
    \underbrace{
    \ln{\left(\frac{P(X_{11})}{P(X_{00})}\right)} - L_{C}^i}_{\eta} +
    \underbrace{ 
    \frac{4\text{Re}[\hat{x}_k(1+i)]}{\sigma_x(1-\mu_x)}}_{\gamma}
    \nonumber\\ & 
    \underbrace{ 
    + \ln{\left(1+\frac{P(X_{10})}{P(X_{11})}e^{-a}\right)}
    + \ln{\left(1+\frac{P(X_{01})}{P(X_{00})}e^{a}\right)}
    }_{\chi}
    \label{eqnn}
\end{align}}
where $a = \frac{4 \text{Re}[\hat{x}_k (1+i)]}{(1-\mu_x)}$. Analyzing this expression, the terms of $\eta$ do not depend on $\boldsymbol{\varphi}$. Also, analyzing $\chi$, we have:
\vspace{-0.0em}
{\begin{align}
    \text{For } a\gg 0  \Longrightarrow \chi \approx -a - \ln{\left(\frac{P(X_{01})}{P(X_{00})}\right)}
    \nonumber\\
    \text{For } a\ll 0  \Longrightarrow \chi \approx -a + \ln{\left(\frac{P(X_{10})}{P(X_{11})}\right)}.
\end{align}
}

It is possible to approximate $\chi$ by a linear function $f(x) = -x + \kappa_1$, where {$\kappa_1 = \ln{\left( \frac{[P(X_{10}) + P(X_{11})]P(X_{00})}{[P(X_{01}) + P(X_{00})]P(X_{11})} \right)}$}. Substituting this approximation in (\ref{eqnn}) leads to:
{\begin{equation}
L_{D}^{i} = \eta + \kappa_1 - \frac{4 \text{Im}[\hat{x}_k (1+i)]}{\sigma_x (1 - \mu_x)}
\end{equation}
}
\vspace{-0.2em}
Using a similar approach to calculate $L_{D}^{ii}$ leads to:
{\begin{equation}
L_{D}^{ii} = \eta + \kappa_2 + \frac{4 \text{Re}[\hat{x}_k (1+i)]}{ \sigma_x(1 - \mu_x)}
\label{Ldi}
\end{equation}
}
\vspace{-0.2em}
To maximize the magnitude of the LLR, the following problem is formulated:
{
\begin{equation}
\boldsymbol{\varphi_k} =
\underset{\boldsymbol{\varphi_k}} \arg \max \sum^{N_{bits}/2} (|L_{D}^{i}| + |L_{D}^{ii}|)
\label{Ldii}
\end{equation}
}
\vspace{-0.1em}
In (\ref{Ldi}) and (\ref{Ldii}), the sole term dependent on $\boldsymbol{\varphi}$ is $\mu_k$. Consequently, as $\mu_k$ resides in the denominator, reducing the magnitude of $(1 - \mu_k)$ is necessary to enhance the LLR. Given that a single $\boldsymbol{\varphi}$ is selected for all users, the mean across all users is taken into account, resulting in the following problem:
{\begin{equation}
\boldsymbol{\varphi} =
\underset{\boldsymbol{\varphi}}\arg \min  \sum^{N_{bits}/2} \sum^{K} |1 - \mu_k|^2
\end{equation}}
Substituting $\mu_k = \mathbf{w_k}^H \mathbf{\bar{h}_k}$ and computing the derivative of the objective function with respect to $\boldsymbol{\varphi}$ and setting the result to zero yields:
\begin{equation}
\boldsymbol{\varphi}_o = \boldsymbol{\beta}^{-1}\boldsymbol{\Psi}.
\label{eq:phi2}
\end{equation}
For ${\sigma_v^2}\approx 0$, this results matches the one obtained using MMSE processing in (\ref{eq:phi}). This solution ensures the maximization of the LLR magnitude, making it suitable for IDD systems.

\subsection{Computational Complexity}

The complexity evaluation of different methods for uplink systems aided by RIS is conducted in terms of floating-point operations (flops) to compute 
$\mathbf{W}$ and $\boldsymbol \varphi$. For the IDD MMSE Active/Passive-RIS (A/P-RIS) algorithm, the primary source of complexity arises from the matrix inversion present in (\ref{eq:phi}) and (\ref{eq:w}), where Cholesky factorization \cite{1494998} is employed to reduce complexity.

Table \ref{tab:1} presents the most significant terms of complexity for each technique. It is evident that techniques using closed-form expressions (CE) have a clear advantage in terms of complexity. However, these techniques often have specific constraints adapted to the determined problem. Most of these methods use iterative processing, and the overall cost depends on the number of iterations, denoted by $I_\mu$\footnote{$I_\mu$ represents the number of iterations required and $\mu\in$ [AO, MMSE, PDD, ADMM, GR, CVX]}.

\begin{table}[!ht]
  \caption{Overall Optimization Complexity}
  \label{tab:1} \vspace{-0.2cm}
\begin{tabular}{|c|c|c|}
\hline
Algorithm & Method  & Optimization Complexity  \\ \hline
{   A/P-RIS} & CE, AO  & $\mathcal{O}[(M^3+N^3)\frac{KI_\text{AO}}{3}]$ \\[3pt] \hline
\cite{9548940} & MM, AO & { \scriptsize $\mathcal{O}[I_\text{AO}(I_\text{cvx}\log(\frac{1}{\epsilon})N^{3.5}(KM^{3.5}+K^2M^{2.5}))]$ }  \\[3pt] \hline
\cite{electronics13050969} & SDR &  $\mathcal{O}[K^{3.5} +(NM)^{3.5} + I_\text{GR}(NM)^{3}]$ \\[3pt] \hline
\cite{9270033} & PDD  & $\mathcal{O}[I_\text{PDD}(K^{1.5}N^{3} + \frac{M^3K}{3})]$ \\[3pt] \hline
\cite{9966649} & CE  & $\mathcal{O}[I_\text{CE}N(3K^{2}M+2K^3)+ \frac{M^3K}{3}]$ \\[3pt] \hline
\cite{9500188} & ADMM  & { \scriptsize $\mathcal{O}[I_\text{MMSE}(M^{2}K + M^3 + I_\text{ADMM}N^3) + \frac{M^3K}{3}]$} \\[3pt] \hline
\end{tabular}
\vspace{0.00cm}

{\scriptsize 
*where Closed-form Expressions (CE), Alternating Optimization (AO), Majorization Minimization (MM), Semidefinite Relaxation (SDR), Penalty Dual Decomposition (PDD), Alternating Direction Method of Multipliers (ADMM).
}
\vspace{-2em}
\end{table}

For a scenario with $[K,M,N]=[12,32,64]$, and assuming that $I_\mu = 5$ for all iterative methods, the A/P-RIS proposed algorithm and the CE \cite{9966649}  exhibit the lowest complexities, with nearly $6 \times 10^6$ complex operations. The ADMM \cite{9500188}  follows with about $7 \times 10^6$ complex operations. The other methods have significantly higher complexities, exceeding $10^{7}$ complex operations. The A/P-RIS algorithm not only ranks among the techniques with the lowest complexity but also provides optimized results for use with IDD techniques.


\section{Numerical Results}
{
We evaluate the proposed IDD scheme through simulations in two uplink scenarios \cite{jpba}: one with only passive RIS and another comparing active and passive RIS. The following schemes are considered:
i) Linear MMSE W/O-RIS, which represents a system that employs a linear MMSE receiver without RIS; ii) Linear MMSE P-RIS, which refers to a system with a linear MMSE receiver and a passive RIS; iii) Linear MMSE A-RIS, which is a system that employs a linear MMSE receiver with an active RIS; iv) IDD MMSE W/O-RIS $\tau$, which is a system that employs a soft MMSE receiver with IDD but without RIS, optimized over $\tau$ iterations; v) IDD MMSE P-RIS $\tau$, which is a system that employs a soft MMSE receiver with  IDD and a passive RIS, optimized over $\tau$ iterations; vi) IDD MMSE A-RIS $\tau$, which is a system that employs a soft MMSE receiver with IDD, combined with an active RIS using $\tau$ iterations.
}


A short-length regular LDPC code \cite{memd,bfpeg,vfap} with a block length of $n=512$ and a rate of $R = 1/2$ is considered with QPSK modulation. The channel is assumed to undergo block fading with perfect and estimated channel state information (CSI) at the receiver. The systems operate at a frequency of 5 GHz, with the direct link weakened due to severe obstruction. To characterize the large-scale fading of the channels, path loss models from the 3GPP standard \cite{access2010further} are utilized: ${\mathrm{PL}{w}}= 41.2 + 28.7\log d$ and ${\mathrm{PL}{s}}=37.3 + 22.0\log d$. The path loss model ${\mathrm{PL}{w}}$ is used for the weak AP-user connection, while ${\mathrm{PL}{s}}$ models the strong connections between AP-RIS and users-RIS channels. To incorporate the effects of small-scale fading, the Rayleigh fading channel model is adopted for all channels.

The AP is positioned at {(0 m, 0 m)}, while user locations are randomly distributed within a 5 m radius circle centered at {(400 m, 0 m)}. The passive RIS is at (400 m, 10 m) and the active RIS is at (200 m, 10 m). The passive RIS location minimizes multiplicative fading \cite{9998527}, while the active RIS balances performance between uplink and downlink scenarios. The system has $K=12$ users, $M=32$ AP antennas and $N=64$ RIS elements. The noise power is set to -100 dBm for $\sigma_s^2$ and 0 for $\sigma_v^2$ in the first scenario, and -95 dBm for both $\sigma_s^2$ and $\sigma_v^2$ (only for active RIS) in the second scenario. The transmit power per user ($P_\text{u}$) is normalized by the code rate, with equal power for all users. For the active RIS scenario, RIS power consumption is limited by the total transmission power ($P_\text{T}$), such that $P_\text{RIS} = 0.1 \times P_\text{T}$ and the power of each user is $P_\text{U} = 0.9 \times P_\text{T} / K$.  Results are presented as total transmission power divided by the number of users ($P_\text{T}/K$).


For CSI estimation, a simplified technique based on \cite{9130088} is employed. Initially, the RIS is turned off to compute the direct links. Then, each user estimates the reflected channels in their respective time slots. Additionally, a Fourier transform-based channel estimation strategy is used, where all RIS elements are active during each time slot, and their reflection coefficients are determined by the Discrete Fourier Transform matrix. For channel estimation, $\sigma_s^2 = -125$ dBm is considered.

\begin{figure}
    \centering
    \includegraphics[width=0.89\linewidth]{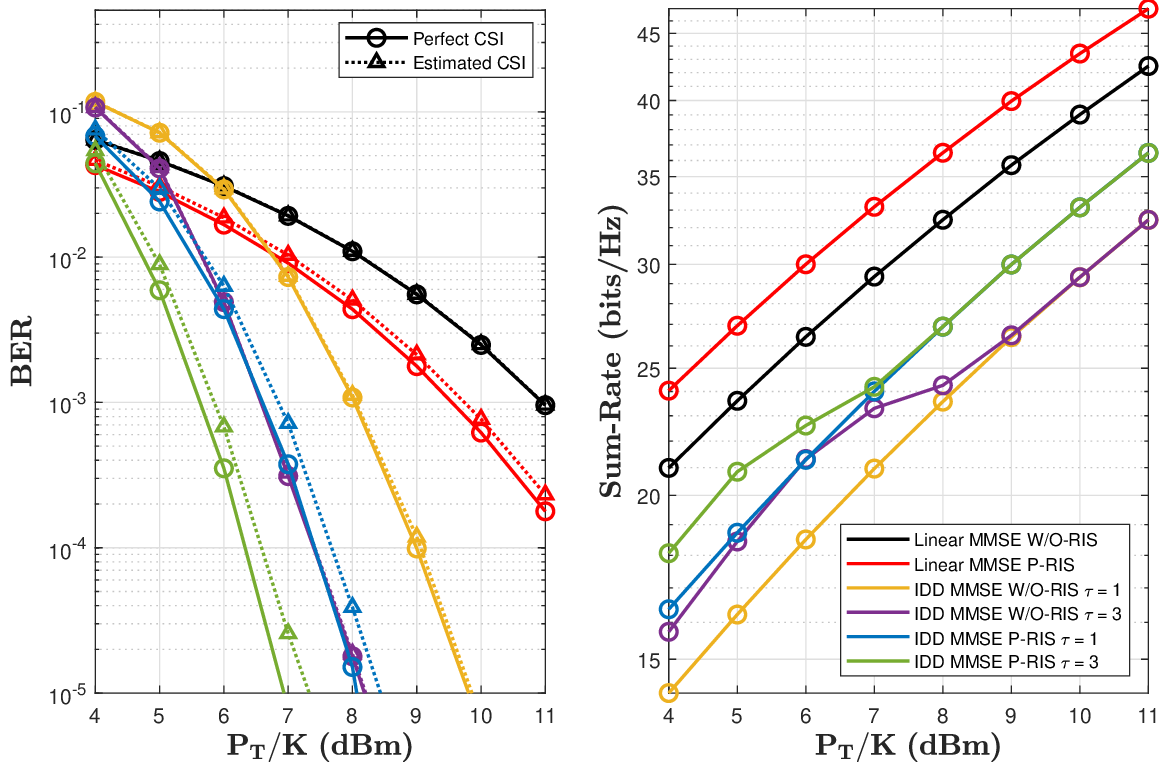}
    \vspace{-0.5em}    
    \caption{Performance for $K = 12$, $M = 32$, $N = 64$, $\sigma_s^2 = -100$ dBm and $\sigma_v^2 = 0$ dBm.}
    \label{r:fig01}
    \vspace{-2em} 
\end{figure}

The BER depicted in Fig. \ref{r:fig01} demonstrates significant improvements for both IDD schemes compared to the {Linear MMSE W/O-RIS}, particularly with a higher number of IDD iterations. Moreover, the proposed {IDD MMSE P-RIS} technique outperforms the {IDD MMSE W/O-RIS} scheme, especially in scenarios with a direct line-of-sight path loss. The proposed scheme with three iterations outperforms the {IDD MMSE W/O-RIS} scheme by approximately 2.7 dB in terms of transmit power per user for the same BER performance. The proposed algorithm has shown resilience against imperfect CSI across different transmit power values.

\begin{figure}
    \includegraphics[width=0.89\linewidth]{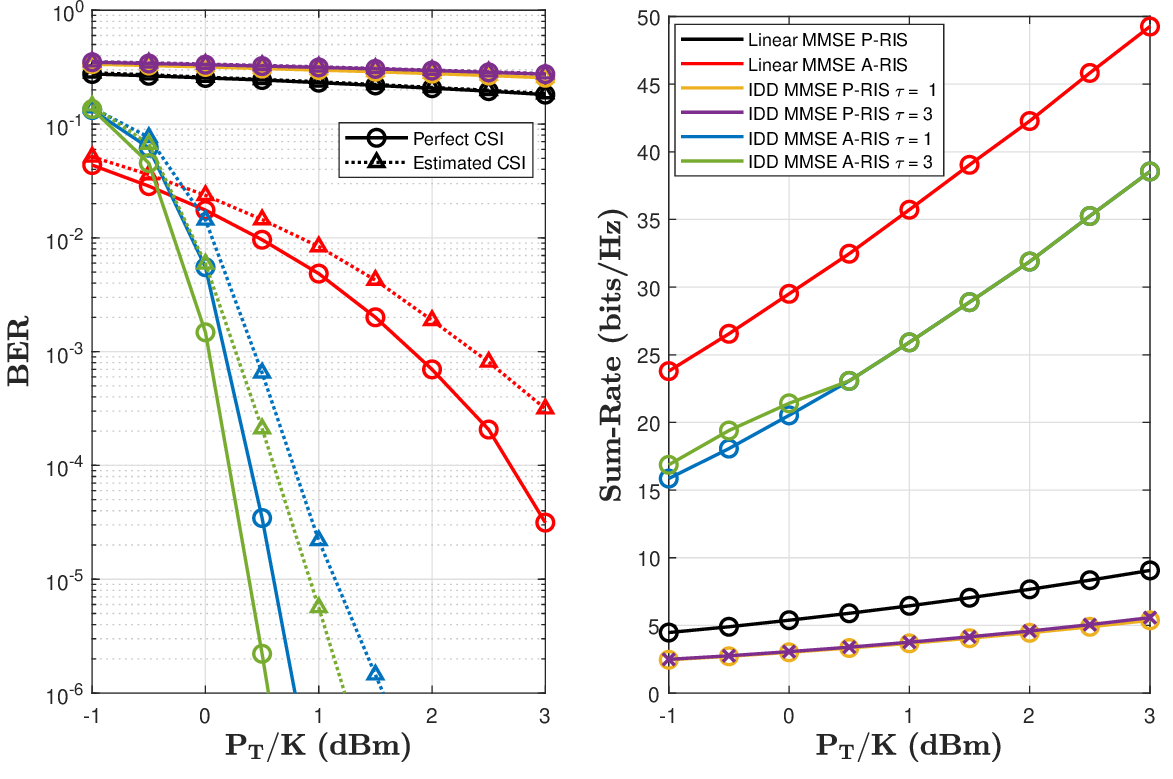}
    \vspace{-0.5em}     
    \caption{Performance for $K = 12$, $M = 32$, $N=64$, $\sigma_s^2 = -95$ dBm and $\sigma_v^2 = -95$ dBm.} 
    \label{fig02}
    \vspace{-2em} 
\end{figure}

Comparisons between an active and a passive RIS are depicted in Fig. \ref{fig02} for a more severe channel ($\sigma_s^2 = -95$ dBm) and lower $P_\text{T}$. It is evident that a passive RIS cannot operate effectively in this system, as it presents high error rates. In contrast, an active RIS shows a significant performance gain, particularly when using an IDD scheme.

Additionally, the sum-rate of users assuming perfect CSI is presented. The SINR at the output of the soft instantaneous MMSE receiver is computed as ${\gamma_{k}^\text{sic}} \triangleq \frac{E[(\hat{x}_{k}^{\text{sic}})^2]}{\text{var}[\hat{x}_{k}^{\text{sic}}]} = \frac{\mu_k^2\sigma_x^2}{\eta_k^2}$. When comparing both schemes, it is noteworthy that there is a performance improvement due to the presence of the RIS.

    \section{Conclusion}

This paper presents an IDD scheme for active and passive RIS-assisted MIMO systems in block-fading channels. Specifically, a closed-form expression for the selection of RIS phases based on the MMSE criterion was formulated, along with a truncation method to satisfy the RIS constraints. Analytical verification confirms that this expression is optimal in terms of LLR refinements within the IDD framework. Simulation results demonstrate a significant BER performance gain after only three iterations. In scenarios with severe LOS path loss, the efficacy of the proposed RIS-IDD algorithm is evident, providing insights for optimizing RIS-assisted MIMO systems.

    \vspace{-1em}
    \bibliographystyle{IEEEtran}
    \bibliography{IEEEabrv,biblio}
\end{document}